\documentclass[a4paper,fleqn,usenatbib]{mnras}

\usepackage{newtxtext,newtxmath}

\usepackage[T1]{fontenc}
\usepackage{ae,aecompl}
\usepackage[dvipsnames]{xcolor}

\usepackage{graphicx}	
\usepackage{amsmath}	

\newcommand{\astrosat}{\textit{AstroSat}}

\newcommand{\pcm}{cm$^{-2}$}

\newcommand{\nh}{$N_{\rm H}$}

\newcommand{\ecps}{erg cm s$^{-1}$~}

\title[GX 339--4]{Spectral Properties of GX~339--4 in the Intermediate State Using AstroSat Observation}
\author[Jana et al.]{
Arghajit Jana$^{1, 2}$\thanks{E-mail: argha0004@gmail.com, argha.jit@mail.udp.cl}, Debjit Chatterjee$^3$, Hsiang-Kuang Chang$^1$, Sachindra Naik$^4$, Santanu Mondal$^3$
\\
$^{1}$Institute of Astronomy, National Tsing Hua University, Hsinchu 300044, Taiwan \\
$^{2}$Instituto de Estudios Astrof\'isicos, Facultad de Ingenier\'ia y Ciencias, Universidad Diego Portales, Av. Ej\'ercito Libertador 441, Santiago, Chile \\
$^{3}$Indian Institute of Astrophysics, II Block Koramangala, Bangalore 560034, India \\
$^{4}$Astronomy \& Astrophysics Division, Physical Research Laboratory, Navrangpura, Ahmedabad, 380 009, India\\
}
\date{Accepted XXX. Received YYY; in original form ZZZ}

\pubyear{2023}

\begin{document}

\label{firstpage}
\pagerange{\pageref{firstpage}--\pageref{lastpage}}
\maketitle

\begin{abstract}
We present the results obtained from the spectral studies of black hole X-ray binary GX~339--4 using \astrosat~ observations during its 2021 outburst. \astrosat~ observed the source in the intermediate state for $\sim600$ ks. The combined spectra of SXT and LAXPC in the $0.7-25$ keV energy range are studied with phenomenological and physical models. The spectral study reveals a receding disc and a contracting corona during the observation period. The outflow rate is found to be increased though the accretion rates did not vary during the observation period. The X-ray flux decreases as the disc recedes and the spectrum becomes hard. At the same time, the Comptonized flux decreases with increasing fraction of thermal emission. This could be plausible that episodic jet ejection modified the corona and reduced Comptonized flux. An iron emission line at 6.4 keV is observed in the spectra of all the orbits of observation. We find that the equivalent width of the iron emission line correlates with the photon index, indicating a decrease in the reflection strength as the spectrum becomes hard. We observe that the disc flux does not follow $F_{\rm DBB}-T^{4}$ relation.
\end{abstract}

\section{Introduction}
\label{sec:intro}
An X-ray spectrum of an outbursting black hole X-ray binary (BHXB) can be approximated by a multi-colour blackbody (MCD) component and a power-law (PL) tail. The MCD emission is believed to be originated in a geometrically thin and optically thick accretion disc \citep{SS73}, while the hard PL tail originates in a Compton corona, located close to the BH \citep[e.g.,][]{HM93,CT95,Done2007}. A fraction of the seed photons get up-scattered with the hot electrons of the Compton corona and produce a power-law tail via inverse-Comptonization \citep{ST80,ST85}. Further, a fraction of the hard photons are reprocessed in the disc and produce an iron K line at $\sim 6.4$~keV and a reflection hump at $\sim 20-40$ keV \citep[][]{Fabian1989,Matt1991}.

An outbursting BHXB commonly shows two major spectral states such as low hard state (LHS) and high soft state (HSS). A BHXB transits from the LHS to HSS or vice-versa through an intermediate state (IMS). A cool disc of temperature $T_{\rm in} \sim 0.1-0.5$ keV, and photon index, $\Gamma \sim 1.5-1.7$ characterizes the X-ray spectrum in the LHS. In the LHS, the hard X-ray flux dominates over the soft X-ray flux. An evolving type-C QPO is observed in the LHS along with a compact and stable jet \citep[e.g.,][]{Nandi2001,Fender2004} which can significantly modify the shape of the spectrum and the accretion geometry. In the HSS, the X-ray spectrum can be characterized by a disc of temperature, $T_{\rm in}\sim 1$ keV and the $\Gamma \geq 2.3$. The soft X-ray flux dominates over the hard X-ray flux in the HSS. No jet or LFQPO is observed in the HSS. In the state transition phase or the IMS, the hard X-ray and soft X-ray fluxes are comparable. The X-ray spectrum is characterized by a disc of temperature $T_{\rm in }\sim 0.5-1$~keV and $\Gamma \sim 2$. A discrete or episodic jet is seen in the IMS. In the IMS, a sporadic type-B and type-A QPOs are observed. One may further divide the IMS as hard-intermediate state (HIMS) and soft-intermediate state (SIMS). The HIMS spectra are generally charactarized by photon index of $\sim 1.8-2$, with the hard X-ray flux dominates. In the SIMS, the soft X-ray flux dominates over the hard X-ray flux, with the photon index $\sim 2-2.2$ and $T_{\rm in}\sim 0.8-1$~keV.

GX\,339-4 was first observed in 1973 \citep{Markert1973ApJ...184L..67M} with the $1-60$ keV MIT X-ray detector onboard OSO-7 satellite. \citet{Parker2016} suggested the mass of the BH as $9.0^{+1.6}_{-1.2} M_{\odot}$, while \citet{Sreehari2019} estimated the mass in the range of $8.3-11.9~M_{\odot}$. The distance of the source is estimated to be $d=8.4\pm0.9$~kpc \citep{Parker2016}. As GX~339-4 is a non-eclipsing binary, the inclination angle should be less than $60^{\rm o}$ \citep{CowleyEtal2002AJ....123.1741C}. \citet{ZdziarskiEtal2019MNRAS.488.1026Z} estimated a lower limit of the inclination as $40^{\rm o}$ from the secondary mass function. Joint modelling of {\it Suzaku}, {\it XMM-Newton}, and {\it NuSTAR} observations was used by \citet{MillerEtal2008ApJ...679L.113M,SM2016} to explain the relativistic broadening of the Fe~K$\alpha$ line and inferred that the source is a highly spinning ($\geq 0.93$) black hole. A distorted Fe~K$\alpha$ line above the continuum is observed in the {\it RXTE} and {\it XMM-Newton} data suggesting a high spin parameter for the black hole \citep{ReisEtal2008MNRAS.387.1489R,PlantEtal2014MNRAS.442.1767P}. Thus, it is evident that the source is extremely spinning, though the exact values of the disc inclination and distance to the source are a topic of intense research.

There are extensive discussions on the origin of temporal, spectral, and multi-wavelength properties of the source during different outbursts using observations with different satellites  \citep{DincerEtal2012ApJ...753...55D,Nandi2012,BachettiEtal2015ApJ...800..109B,BasakZdzi2016MNRAS.458.2199B}. Most of the works reached at the conclusion that the spectra of GX\,339-4 can be well explained by considering a cold thermal disc, hot corona, and a reflection component. In addition, a Gaussian component is needed to fit the Fe~K$\alpha$ line, which is believed to be originated from reflection. However, this source showed a positive correlation between the equivalent width (EW) and mass accretion or photon index ($\Gamma$) \citep[see][]{Tomsicketal2009,DebnathEtal2015MNRAS.447.1984D}, which is somewhat opposite from what numerical simulation predicts \citep{ZyckiBozena1994,Garcia2013,MondalEtal2021MNRAS.505.1071M}. Such positive correlation can be observed if the disc moves inward (increasing $\Gamma$), thereby increasing the number of hard photons intercepted in the disc, which increases the reflection; hence, the EW increases \citep[e.g.,][]{Zdziarski1999}.

GX~339--4 showed a long X-ray outburst in 2021, which lasted for about 10 months \citep{Pal2021Atel,Corbel2021ATel}. The outburst was extensively studied with several X-ray observatories, such as {\it Swift} \citep{Pal2021Atel}, {\it NuSTAR} \citep{Garcia2021ATel}, {\it NICER} \citep{Wang2021ATel}, Insight-HXMT \citep{Liu2021ATel}, {\it INTEGRAL} \citep{Ferrigno2021ATel} and {\it AstroSat} \citep{Husain2021ATel}. The source was also studied in the optical \citep{Saikia2021ATel} and radio wavebands \citep{Husain2021ATel}. Using Insight-HXMT, \citet{Liu2022} studied the source during its hard-to-soft state transition. \citet{Stiele2023} studied the evolution of the type-B QPO using the {\it NICER} observation. \citet{Yang2023} investigated the type-B QPO and fast variability of the source during the state transition using Insight-HXMT observation. Using the time-averaged spectra of {\it AstroSat} and {\it NICER}, \citet{Peirano2023} suggested of a dual corona in the source.

In general, the accretion dynamics is well studied in the LHS and HSS. However, it is not well understood in the IMS \citep[e.g.,][]{AJ2022c}. It is believed that the accretion geometry changes significantly during the IMS \citep[e.g.,][]{Gardner2013,Yang2015,AJ2022a}. The presence of the discrete ejection also makes the IMS complex. To understand the IMS better, we studied the well-known BHXB GX~339--4 in the IMS during its 2021 outburst in this paper. \astrosat~ observed GX~339--4 in the IMS between 30 March 2021 and 6 April 2021 for an exposure of 600~ks. Although, \citet{Peirano2023} studied the same {\it AstroSat} observation, they studied the spectrum only obtained on 30 March 2021. Here, we used all the data obtained from the entire {\it AstroSat} observation from 30 March 2021 to 6 April 2021. The timing properties of the source are presented in \citet{MondalEtal2023arXiv230303742M}. In this paper, we present the results obtained from the spectral analysis of the source in the IMS. The paper is organized in the following way. In \S2, we present the observations and data extraction processes. The analysis and results are presented in \S3. In \S4, we discuss our findings. Finally, in \S5, we summarize our findings.

\section{Observation and Data Reduction}
\label{sec:obs}
\astrosat~ is the first Indian multi-wavelength astronomical satellite that provides a broad-band coverage from optical to hard X-ray bands for exploring the nature of the cosmic sources \citep{Agrawal2006}. It consists of five sets of instruments, namely Soft X-ray Telescope \citep[SXT;][]{Singh2017}, Large Area X-ray Proportional Counters \citep[LAXPC;][]{Antia2017}, Cadmium Zinc Telluride Imager \citep[CZTI;][]{Rao2017}, a Scanning Sky Monitor \citep{Ramadevi2018}, and Ultraviolet Imaging Telescope \citep[UVIT;][]{Tandon2017}, onboard the satellite. In the present work, we used data from the SXT and LAXPC.

The SXT is a soft X-ray focusing telescope that works in $0.3-8$ keV energy range. The effective area and energy resolution of the instrument are 128 cm$^2$ and $5-6$ percent at 1.5 keV and 22 cm$^2$~ and 2.5 percent at 6 keV, respectively. GX~339-4 was observed with SXT in the photon counting (PC) mode, at a time resolution of 2.4 s. The level-1 data were reprocessed using standard SXT pipeline software {\tt AS1SXTLevel2-1.4b1} to obtain cleaned event files from each orbit of observation. As the source was bright in soft X-rays (count rate of $>$ 40 count/sec), the SXT data are affected due to photon pile-up. To avoid the pile-up effect, we chose an annular region with a fixed outer radius of 10 arcmin and a variable inner radius. The pile-up was checked from the spectral distortion. We found that the pile-up effect was removed when the inner radius of 8 arcmin is considered in our analysis. We used background spectra and response matrix files that are supplied by the SXT instrument team. Using {\tt SXTARFMODULE} tools, the auxiliary response files (ARF) are generated. In the current work, we used SXT spectra in the $0.7-7.0$~keV range.

\astrosat~ has three LAXPC units that are sensitive to the X-ray photons in $3-80$ keV energy range, with a total effective area of 8000 cm$^2$ at 15 keV. The timing and spectral resolutions of the LAXPC are 10 $\mu s$ and 12 percent at 22 keV, respectively. In our analysis, we used only event mode data from LAXPC20 unit. Due to high background and gain issues with the instruments, we did not use data from LAXPC10 and LAXPC30 \citep{Antia2017,Antia2021}. We extracted spectra, background spectra, and response matrices using standard data analysis tools {\tt LAXPCsoftware} (LAXPCSOFT; version 2022 August 15). We used LAXPC spectra in the $3-25$~keV energy range in our analysis. We did not use spectra above 25 keV as the data are background dominated above 25~keV.

\begin{figure}
\centering
\includegraphics[width=8.5cm]{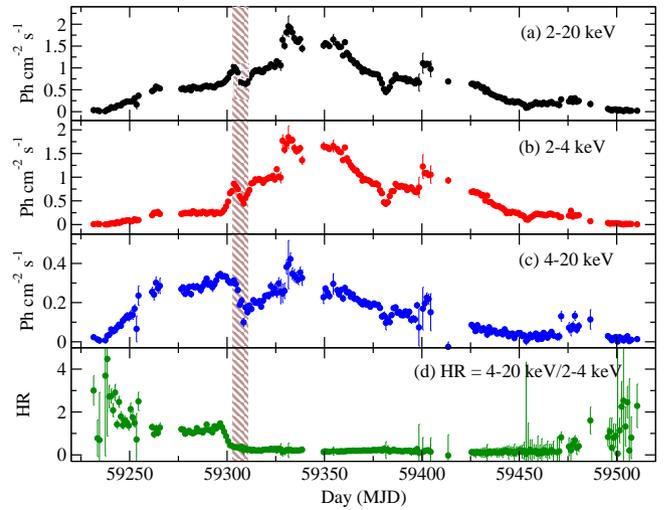}
\caption{Outburst profile of GX~339--4 during the 2021 outburst. The variation of (a) $2-20$~keV count rate, (b) $2-4$~keV count rate, (c) $4-20$~keV count rate, and (d) hardness ratio (count rate in $4-20$ keV range / count rate in $2-4$ keV range) are shown. The light curves are obtained from MAXI/GSC. The shaded grey region marks the duration of the {\it AstroSat} observation of the source.}
\label{fig:lc}
\end{figure}

\begin{figure*}
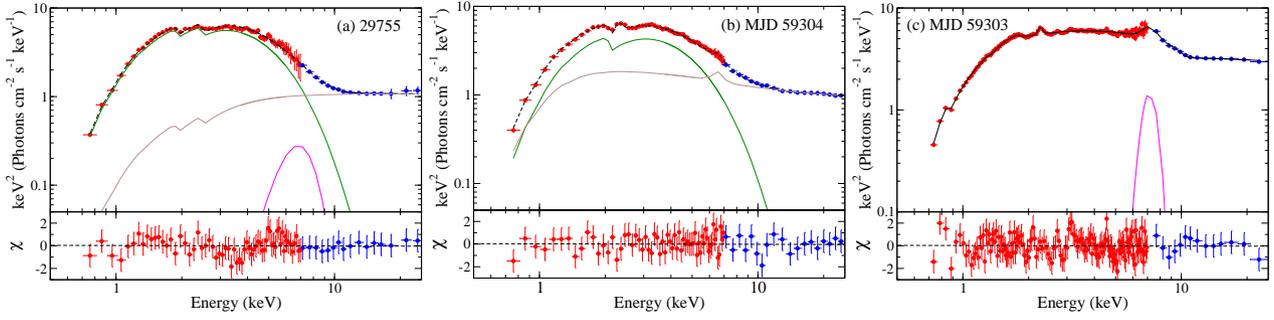

\centering
\includegraphics[width=5.5cm]{755.eps}
\includegraphics[width=5.5cm]{304.eps}
\includegraphics[width=5.5cm]{303-tcaf.eps}
\caption{Left panel : Representative spectrum of GX~339--4 for orbit number 29755 (MJD~59303.12), fitted with Model-1. The red and blue points represent the SXT and LAXPC data, respectively. The black dashed, grey solid, dark green solid, and magenta lines represent the total emission, Comptonized emission, disc-blackbody and iron line emission, respectively. Middle panel : Model-2 fitted spectrum for MJD~59304. The black dashed, grey solid and dark green solid lines represent the total emission, Comptonized \& reprocessed emission, and disc-blackbody, respectively. Right panel : Spectrum for MJD~59303, fitted with Model-3. The black and magenta lines represent the total and iron line emission, respectively.}
\label{fig:spec}
\end{figure*}

\begin{figure}
\centering
\includegraphics[width=8.5cm]{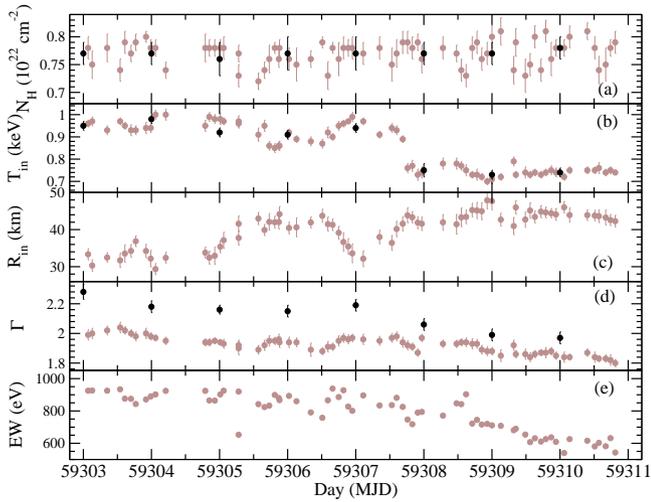}
\caption{Variation of (a) line-of-sight equivalent hydrogen column density ($N_{\rm H}$) in $10^{22}$ cm$^{-2}$, (b) inner disc temperature ($T_{\rm in}$) in keV, (c) inner disc radius ($R_{\rm in}$) in km, (d) photon index ($\Gamma$), and (e) equivalent width (EW) of Fe~K$\alpha$ line in eV are shown during our observation period. The black and grey points represent the result of one day-averaged and orbit wise spectral fitting with Model-2 and Model-1, respectively.}
\label{fig:spec-par}
\end{figure}

\begin{figure}
\centering
\includegraphics[width=8.5cm]{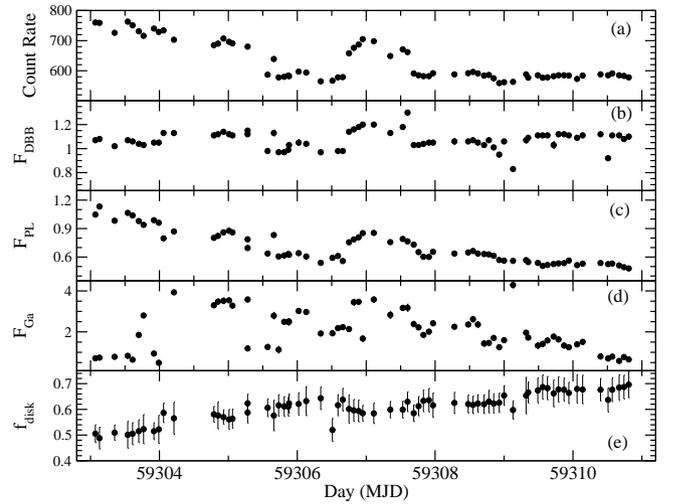}
\caption{Evolution of (a) $3-25$~keV {\it AstroSat}/LAXPC count rate in count s$^{-1}$, (b) thermal disc flux ($F_{\rm DBB}$) in $10^{-8}$ ergs cm$^{-2}$ s$^{-1}$, (c) Comptonized power-law flux ($F_{\rm PL}$) in $10^{-8}$ ergs cm$^{-2}$ s$^{-1}$, (d) iron line flux ($F_{\rm Ga}$) in $10^{-10}$ ergs cm$^{-2}$ s$^{-1}$, and (e) thermal emission fraction ($f_{\rm disc}$) are shown during our observation period.}
\label{fig:flux}
\end{figure}

\section{Analysis and Results}
\label{sec:result}

\subsection{Outburst Profile}
\label{sec:prof}
GX~339--4 showed an outburst in the first week of January 2021 that lasted for about 10 months. The 2021 outburst was extensively monitored in multi-wavelength bands. We show the outburst profile of GX~339--4 during the 2021 outburst in Figure~\ref{fig:lc}. In the top three panels, we show the variation of MAXI/GSC light curves in (a) $2-20$~keV, (b) $2-4$~keV, and (c) $4-20$~keV energy ranges, respectively. In the bottom panel of Figure~\ref{fig:lc}, we show the variation of hardness ratio (HR). The HR is defined as the ratio of the count rate in the $4-20$ keV range to the count rate in the $2-4$~keV range.

At the beginning of the outburst, the count rate increases rapidly in all energy bands till 18 February 2021 (MJD~59263). Later, the count rate increased slowly till 20 March 2021 (MJD~59293). Then, the soft X-ray ($2-4$ keV energy band) count rate increased rapidly till 30 March 2021 (MJD~59303), although the hard X-ray count rate did not change. As a result, the HR decreased sharply as the source entered the IMS. Then, the count rate in all energy bands decreased sharply, showing an alternative flux profile (AFP) similar to the `flip-flop' profile (FFP) nature of the light curve between 30 March 2021 (MJD~59303) and 10 April 2021 \citep[MJD~59314; see][for details]{MondalEtal2023arXiv230303742M}. The AFP is also seen in a shorter timescale on MJD~59305--59306. Since MJD~59308, the soft X-ray count rate increased rapidly compared to the hard X-ray band. On 4 April 2021 (MJD~59331), the X-ray flux attained its peak as the source entered the HSS.

After 4 April 2021, the source entered the declining phase of the outburst as both soft X-ray and hard X-ray count rates decreased slowly till 3 July 2021 (MJD~59398). Another AFP was observed in the declining phase between 12 June 2021 (MJD~59377) and 21 June 2021 (MJD~59386). A sudden increase in the X-ray count rate was observed after 3 July 2021. Following that, the X-ray intensity slowly decreased till 13 September 2021 (MJD~59470). The HR was observed to be constant till that day. Thereafter, the HR increased, indicating the source moved to the IMS. After 29 September 2021 (MJD~59486), the source was observed to be in the LHS as the HR increased and the X-ray count rate decreased further. GX~339--4 remained in the LHS till the end of the outburst.

\subsection{Spectral Analysis}
\label{sec:spec-analysis}
We used combined $0.7-7$~keV SXT and $3-25$~keV LAXPC data for both orbit-wise and day-wise spectral study. We carried out the spectral analysis in $0.7-25$~keV energy range in HEASEARC’s spectral analysis software package {\tt XSPEC} v12.10 \citep{Arnaud1996}. We used both phenomenological and physical models to study the spectral properties of the source. The orbit-wise spectra are studied with phenomenological models, while the day-wise spectra are studied with the physical models. For the phenomenological model, we used the simple \textsc{diskbb} \citep{Mitsuda1984,Makishima1986} plus \textsc{nthcomp} model \citep{Z96,Zycki1999} to fit the spectrum. For the physical model, we used the relativistic reflection model \textsc{relxillLp} \citep{Garcia2013,Dauser2016}, and \textsc{JeTCAF} \citep{CT95,MondalChakrabarti2021ApJ920} model in our fitting.

\subsubsection{Phenomenological Model:}
\label{sec:pheno}
We started our analysis with the phenomenological two-component model (hereafter Model-1): thermal multi-colour blackbody emission and Comptonized emission for the spectral fitting. The thermal and non-thermal emissions were modelled by \textsc{diskbb} and \textsc{nthcomp} in {\tt XSPEC}, respectively. Along with these, we also used \textsc{tbabs} and a \textsc{Gaussian} function for the Galactic absorption and iron K$\alpha$ emission line, respectively. We fixed the Gaussian line energy at 6.4 keV. During the fitting, we tied the seed photon temperature ($kT_{\rm bb}$) of \textsc{nthcomp} with the inner disc temperature ($T_{\rm in}$) of the \textsc{diskbb} model. Initially, we kept the hot electron temperature ($kT_{\rm e}$) free. However, we could not constrain it. Thus, we kept the $kT_{\rm e}$ frozen at 50~keV during the spectral analysis. We used \textsc{WILM} abundances \citep{Wilms2000} and cross-section of \citet{Verner1996} in our analysis. We added a systematic of 3\% while fitting the data \citep[e.g.,][]{Antia2021}. We also applied a gain correction to the SXT spectra with a fixed slope of 1 to flatten the residuals at 1.8 and 2.2 keV using \texttt{gain fit} command in {\tt XSPEC.}

From the \textsc{diskbb} normalization ($N_{\rm DBB}$), we calculated the inner disc radius ($R_{\rm in}$). The $R_{\rm in}$ is related to the $N_{\rm DBB}$ as $N_{\rm DBB}=(r_{\rm in}/D_{\rm 10}^2) \cos{\theta}$, where $r_{\rm in}$ is the apparent disc radius in km, $D_{\rm 10}$ is the source distance in the unit of 10 kpc, and $\theta$ is the inclination angle. The inner radius of the disc ($R_{\rm in}$) is given by, $R_{\rm in} = \xi \kappa^2 r_{\rm in}$, where $\xi = 0.41$ is the correction factor \citep{Kubota1998}, and $\kappa=1.7-2.0$ is the spectral hardening factor \citep{Shimura-Takahara1995}. Here, we calculated $R_{\rm in}$, considering $\kappa=1.8$, $\theta=30$ degrees, and $d=8.4\pm0.9$~kpc \citep{Parker2016}. For the iron emission line, we calculated the equivalent width (EW) in {\tt XSPEC} using \textsc{eqw} command. Using \textsc{cflux}, we calculated absorption corrected flux for each of the three components. We calculated the disc flux ($F_{\rm DBB}$), iron line flux ($F_{\rm Ga}$), and Comptonized flux ($F_{\rm PL}$) in the range of $0.001-10$~keV, $0.1-10$~keV and $0.1-500$~keV, respectively. We also considered a high energy cutoff at 200~keV while calculating $F_{\rm PL}$. The total flux ($F_{\rm tot}$) is calculated as $F_{\rm tot}=F_{\rm DBB}+F_{\rm PL}+F_{\rm Ga}$. We also estimated the fraction of thermal emission as $f_{\rm disc}=F_{\rm DBB}/F_{\rm tot}$. The detailed spectral analysis results are tabulated in Table~\ref{tab:spec-res}.

We obtained good fits for all orbits-wise spectra with the Model-1. The left panel of Figure~\ref{fig:spec} shows the best-fitted spectra for orbit 29755 (MJD~59303.12), fitted with Model-1. The residual is shown in the bottom panel. The red and blue points represent the SXT and LAXPC data, respectively. The black dashed, grey solid, dark green solid, and magenta lines represent the total emission, Comptonized emission (\textsc{nthcomp}), disc-blackbody (\textsc{diskbb}) and iron emission line (\textsc{gaussian}), respectively. We show the variation of (a) line-of-sight hydrogen column density ($N_{\rm H}$) in $10^{22}$ cm$^{-2}$, (b) inner disc temperature ($T_{\rm in}$) in keV, (c) inner disc radius ($R_{\rm in}$) in km, (d) photon index ($\Gamma$), and (e) equivalent width (EW) of iron K$\alpha$ line in eV in Figure~\ref{fig:spec-par}. {The grey points represent the spectral analysis result obtained with the Model-1 in all panels}. Figure~\ref{fig:flux} shows the variation of (a) $3-25$ keV AstroSat/LAXPC count rate in count s$^{-1}$, (b) thermal disc flux ($F_{\rm DBB}$) in 10$^{-8}$ ergs cm$^{-2}$ s$^{-1}$, (c) Comptonized power-law flux ($F_{\rm PL}$) in 10$^{-8}$ ergs cm$^{-2}$ s$^{-1}$, (d) iron line flux ($F_{\rm Ga}$) 10$^{-8}$ ergs cm$^{-2}$ s$^{-1}$, and (e) thermal emission fraction ($f_{\rm disc}$).

From the spectral analysis with Model-1, we observed that $N_{\rm H}$ varied in the range of $0.7-0.8 \times 10^{22}$ \pcm~ during the observation duration. The $T_{\rm in}$ was found to fluctuate in the range of $0.9-1$~keV until the \astrosat~ orbit 29819 (MJD~59307.59). Then, the $T_{\rm in}$ decreased to $\sim 0.8$~keV. The $T_{\rm in}$ was observed to be in the range of $\sim 0.7-0.8$~keV for the rest of the observations. We also observed a variation in $R_{\rm in}$. At the beginning of the observation, $R_{\rm in}$ was observed to be $\sim 32$ km. Later, $R_{\rm in}$ was found to increase with $R_{\rm in}\sim 40-45$~km, indicating that the disc is moving outward. 

At the beginning of the observations, the photon index $\Gamma$ is found to be $\sim 2$. Later, it decreased to $\Gamma \sim 1.8$, as the source count rate decreased. The EW of the iron emission line was found to decrease as $\Gamma$ decreased. The Pearson correlation between the $\Gamma$ and EW is found to be $r=0.84$ with $p$-value of $<10^{-3}$. Such a positive correlation for this source is consistent with the findings in the literature. As the disc is receding away from the BH in progressive orbits of AstroSat observation, the gravitational effects decrease. Therefore, the EW is decreasing. As the EW represents the reflection, the reflection strength decreases as the spectrum becomes hard. At the same time, we observed the fraction of the thermal emission ($f_{\rm disc}$) increased towards the end of the observation period.

\begin{figure*}
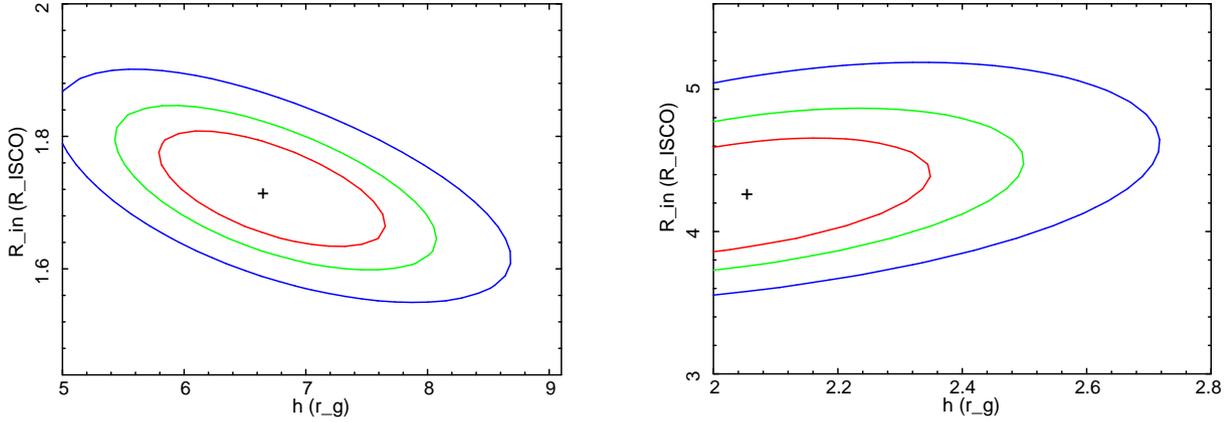

    \centering
    \includegraphics[angle=270,width=8.5cm]{fig5a.eps}
    \includegraphics[angle=270,width=8.5cm]{fig5b.eps}
    \caption{2D-contours between the the corona height ($h$) and inner disc radius ($R_{\rm in}$) for the AstroSat observations on MJD 59303 in the left panel, and for MJD 59310 in the right panel.}
    \label{fig:h}
\end{figure*}

\subsubsection{RelxillLp:}
\label{sec:relxill}
Next, we fitted the one-day-averaged spectra in $0.7-25$~keV energy range with the physical \textsc{relxillLp} model, which replaces the \textsc{nthcomp} in the phenomenological model. The \textsc{relxillLp} model considers the lamp-post geometry. In this model, the corona is considered to be a point source, located at a height of `h' from the BH. The hard X-ray photons from the corona are reprocessed in the disc and produce Fe~K-line and reflection hump in the observed spectrum. The complete model (hereafter Model-2) reads in {\tt XSPEC} as \textsc{TBabs*(diskbb+relxillLp}). During the analysis, we kept the spin parameter frozen at 0.998, the inclination angle at 30 degrees, and the outer disc radius at 400 $R_{\rm g}$. As we could not constrain the cutoff energy, we also fixed it at $300$~keV. The fitting with the \textsc{relxillLp} model gave us an acceptable fit ($\chi^2/$dof $\sim 1$) for all the one-day-averaged spectra. The spectral analysis result is tabulated in Table~\ref{tab:relxill}.

From the day-wise spectral analysis with Model-2, we obtained a similar result as with Model-1. In Figure~\ref{fig:spec-par}, we show the variation of (a) line-of-sight hydrogen column density ($N_{\rm H}$) in $10^{22}$ cm$^{-2}$, (b) inner disc temperature ($T_{\rm in}$) in keV, and (d) photon index ($\Gamma$) obtained from Model-2 spectral fits with black points. During our observation period, the $T_{\rm in}$ is found to decrease from $\sim 1$~keV to $\sim 0.7$~keV. The $\Gamma$ from the \textsc{relxillLp} model is somehow softer than the findings with the \textsc{nthcomp} model. This is because the \textsc{relxillLp} includes reprocessed emission, which is absent in the phenomenological model fitting. We found that the ionization parameter ($\xi$) and the iron abundances ($A_{\rm Fe}$) remain same throughout the observation period, with $\log \xi \sim 4$ and $A_{\rm Fe} \sim 4~A_{\sun}$. The height of the corona is observed to change from $\sim 6.5~R_{\rm g}$ to $\sim 2~R_{\rm g}$ during the observation period. The inner disc radius is found to increase from $\sim 1.7$ $R_{\rm ISCO} $ to $\sim 4.3$ $R_{\rm ISCO} $, indicating a receding disc.

The middle panel of Figure~\ref{fig:spec} shows the Model-2 fitted spectrum for MJD 59304. The black dashed, grey solid, and dark green solid lines represent the total emission, Comptonized \& reprocessed emission (\textsc{relxillLp}) and disc-blackbody (\textsc{diskbb}) emission, respectively. The variation of the spectral parameters obtained with Model-2 is shown with black points in Figure~\ref{fig:spec-par}. We show the variation of (a) the line-of-sight hydrogen column density ($N_{\rm H}$) in $10^{22}$ cm$^{-2}$, (b) inner disc temperature ($T_{\rm in}$) in keV, and (d) photon index ($\Gamma$). Figure~\ref{fig:h} shows the 2D-contours between the height of the corona ($h$) and the inner disc radius ($R_{\rm in}$) for the AstroSat observations on MJD 59303 and MJD 59310 in the left and right panels, respectively. In Figure~\ref{fig:spec2}, we show the variation of (d) coronal height ($h$ in $R_g$) and (e) inner disc radius ($R_{\rm in}$ in $R_{\rm ISCO}$) with MJDs with red points.

\subsubsection{JeTCAF:}
\label{sec:tcaf}
We also used physical model \textsc{JeTCAF}, which includes jet/outflows \citep{MondalChakrabarti2021ApJ920} for further spectral analysis. In this model, the accretion flow has two components: an optically thick, geometrically thin high viscous Keplerian flow on the equatorial plane submerged inside an optically thin, low viscous sub-Keplerian flow. The Keplerian flow produces the soft thermal multi-color disc blackbody emission. The sub-Keplerian flow forms an axis-symmetric shock which produces the dynamic corona at the post-shock region. The corona up-scatters the soft photons from the Keplerian flow and produces the hard power-law emission. The corona also acts as the base of the jet, i.e., the jet is launched from the corona \citep[see ][for details]{CT95,skc1999,skc2018,MondalChakrabarti2021ApJ920}.

The \textsc{JeTACF} model has six free parameters, such as the mass of the black hole ($M_{\rm BH}$), the Keplerian disk accretion rate ($\dot{m}_{\rm d}$ in Eddington rate  $\dot{m}_{\rm Edd}$), the sub-Keplerian halo rate ( $\dot{m}_{\rm h}$ in  $\dot{m}_{\rm Edd}$), the size of the corona or shock location ($X_s$ in $R_g$), shock compression ratio ($R$, ratio of the matter density in post-shock to pre-shock region), and collimation factor of the jet/outflow ($f_{\rm col}=\theta_{\rm out}/\theta_{\rm in}$; $\theta_{\rm out}$ and $\theta_{\rm in}$ are angle subtended by the outflow and inflow, respectively). 

The \textsc{JeTCAF} model does not contain the iron emission line. Hence, we added a Gaussian line at 6.4~keV to incorporate the iron line during the spectral analysis. The spectral model (hereafter Model-3) reads in \texttt{XSPEC} as \textsc{TBabs*(JeTCAF + Gaussian)}. We obtained an acceptable fit with the \textsc{JeTCAF} model for all the one-day-averaged spectra. The $\dot{m}_{\rm d}$ and $\dot{m}_{\rm h}$ varied in the range of $\sim 1.7-2.1$ $\dot{m}_{\rm Edd}$ and $\sim 0.2-0.3$ $\dot{m}_{\rm Edd}$, respectively. The size of the corona was $\sim 75$ $R_g$ at the beginning of the outburst, and then it decreased to $\sim 50$ $R_g$.

We also calculated the mass outflow rate ($\dot{m}_{\rm out}$) from the $R$ and $f_{\rm col}$. The ratio of outflow to inflow rate is given by $R_{\dot{m}}=f_{\rm col} f_0^{3/2}\frac{R}{4} \exp[\frac{3}{2}-f_0]$, where $f_0=R^2/(R-1)$ \citep{skc1999}. We found that $\dot{m}_{\rm out}$ increased to $0.33\pm0.07$ $\dot{m}_{\rm Edd}$ on MJD 59305, from $0.21\pm0.02$ $\dot{m}_{\rm Edd}$ on MJD 59303. Later, it decreased briefly, and then $\dot{m}_{\rm out}$ increased to $0.54\pm0.13$ $\dot{m}_{\rm Edd}$ on MJD 59310.

\begin{figure}
\centering
\includegraphics[width=8.5cm]{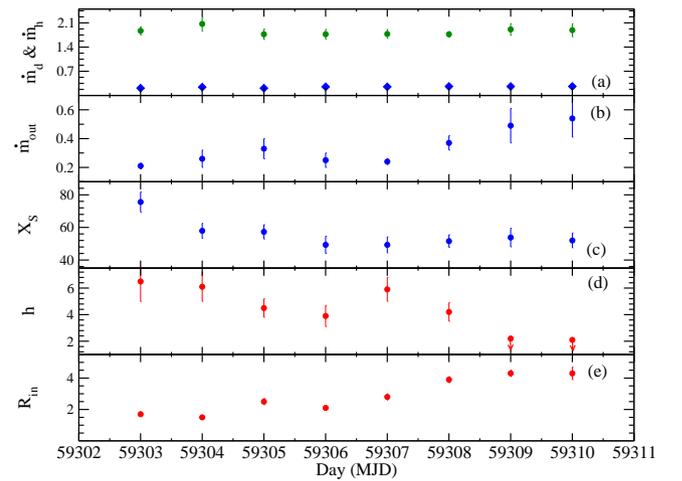}
\caption{Evolution of (a) Keplerian disc accretion rate ($\dot{m}_{\rm d}$ in $\dot{m}_{\rm Edd}$; \textcolor{OliveGreen}{green}) \& sub-Keplerian flow rate ($\dot{m}_{\rm h}$ in $\dot{m}_{\rm Edd}$; \textcolor{blue}{blue}), (b) mass outflow rate ($\dot{m}_{\rm out}$ in $\dot{m}_{\rm Edd}$) from Model-3, and (c) the size of the corona ($X_s$ in $R_g$), (d) coronal height ($h$ in $R_g$) and (e) inner disc radius ($R_{\rm in}$ in $R_{\rm ISCO}$) from Model-2.}
\label{fig:spec2}
\end{figure}

From the spectral modelling with Model-3, we obtained the mass of the BH as $7.3\pm0.6$ $M_{\odot}$. The model normalization is also found to be constant across all the observations at $N = 11.5\pm1.5$, as expected from theoretical point-of-view. In the right panel of Figure~\ref{fig:spec}, we show the Model-3 fitted spectrum for MJD 59303. The black and magenta lines represent the total and iron emission line (\textsc{gaussian}), respectively. In Figure~\ref{fig:spec2}, we show the variation of (a) Keplerian disc accretion rate ($\dot{m}_{\rm d}$ in $\dot{m}_{\rm Edd}$) \& sub-Keplerian flow rate ($\dot{m}_{\rm h}$ in $\dot{m}_{\rm Edd}$), (b) mass outflow rate ($\dot{m}_{\rm out}$ in $\dot{m}_{\rm Edd}$), and (c) size of the corona ($X_s$ in $R_g$). The spectral analysis result is tabulated in Table~\ref{tab:tcaf}.

\begin{figure}
\centering
\includegraphics[width=8.5cm]{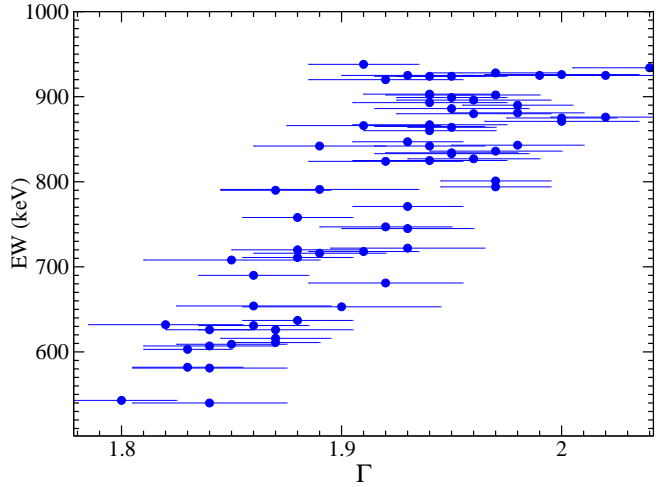}
\caption{Variation of equivalent width (EW) of Fe K$\alpha$ line as a function of power-law photon index ($\Gamma$).}
\label{fig:eq}
\end{figure}

\section{Discussion}
\label{sec:dis}
We studied the spectral and timing properties of GX~339-4 in $0.7-25$~keV energy range, using \astrosat~ observations during its 2021 outburst. \astrosat~ observations were carried  in the rising phase of the outburst when the source was in the intermediate state. We used LAXPC data in $3-25$~keV range for the timing study, while we used combined SXT and LAXPC data in $0.7-25$~keV range for the spectral study. 

\subsection{Comparison between different spectral models}
\label{sec:comparison}

During our spectral analysis, we obtained \nh~ in the range of $\sim0.7-0.8 \times 10^{22}$ \pcm~ from all the models. Our estimated value of \nh~ is greater than the values reported in earlier studies, where \nh~ was found to be $\sim 0.5-0.6 \times 10^{22}$ \pcm~ \citep{Motta2009}. In the previous studies, the abundances were considered to be \textsc{AGNR} \citep{Anders1989}, whereas we considered \textsc{WILMS} abundances in this work. We fitted the data considering \textsc{ANGR} abundances, where we found that \nh~ varied in the range of $\sim 0.5-0.6 \times 10^{22}$ \pcm.

We found a softer $\Gamma$ while fitting the data with Model-2, compared to the Model-1. From the Model-1 fitting, the $\Gamma$ varied in the range of $\Gamma \sim 1.8-2$, while in the case of the Model-2 fitting, $\Gamma$ was found to be in the range of $2-2.3$. As the reprocessed emission is not included in Model-1, we obtained a somehow harder $\Gamma$. Another possibility for different $\Gamma$ is using different primary continuum models in Model-1 and Model-2. In Model-1, \textsc{nthcomp} is used as the primary continuum, while \textsc{cutoffpl} is used in Model-2 as the continuum model. The \textsc{nthcomp} and \textsc{cutoffpl} treat the low-energy part of the spectrum differently, which may result in different $\Gamma$ \citep[e.g.,][]{AJ2021maxij0637}. However, the difference will not be large as the $\Gamma$ is calculated from the high energy part of the spectrum. We tested this by replacing \textsc{nthcomp} with \textsc{cutoffpl} in Model-1, and \textsc{relxillLpCp} with \textsc{RelxillLp} in Model-2, and we found $\Delta \Gamma \sim 0.01 \pm 0.02$, in both the cases. Nonetheless, a different $\Gamma$ is obtained from both models, the variation of the $\Gamma$ is similar. The inner disc temperature ($T_{\rm in}$) is obtained from both Model-1 and Model-2. We obtained a similar variation of the $T_{\rm in}$ from both models. At the beginning of the \astrosat~ observation, the $T_{\rm in}$ was $\sim 1$~keV. Later, the $T_{\rm in}$ decreased to $\sim 0.7$~keV.

We calculated the inner disc radius ($R_{\rm in}$) from the \textsc{diskbb} normalization in Model-1. We observed that $R_{\rm in}$ increased from $\sim 30$~km to $\sim 45$~km during the observation period. The spectral analysis with Model-2 found that the $R_{\rm in}$ increased from $\sim 1.7$ $R_{\rm ISCO}$ to $\sim 4.3$ $R_{\rm ISCO}$ during the observation period. The \textsc{diskbb} yields somehow a higher value of $R_{\rm in}$ compared to the \textsc{relxillLp} model. This is because the \textsc{diskbb} does not consider the spin of the black hole. Nonetheless, both models indicated a receding disc.

We obtained the coronal properties from Model-2 \& Model-3. In Model-2, the \textsc{relxillLp} considers a lamp-post geometry, with the point source located at a distance of `h' from the BH. We observed that `h' decreased from $\sim 7$~$R_g$ to $\sim 2$ $R_g$ during the observation period, implying a contracting corona. On the other hand, in Model-3, \textsc{JeTCAF} considers an extended corona, with the disc truncated at the boundary of the corona. The spectral analysis with Model-3 revealed a contracting corona with its size decreasing from $\sim 75$ $R_g$ to $\sim 50$ $R_g$ during the AstroSat observation period. As the \textsc{relxillLp} and \textsc{JeTCAF} models consider different geometries of the corona, we obtained different sizes of the corona. However, both models suggested a contracting corona.


\begin{figure}
    \centering
    \includegraphics[width=8.5cm]{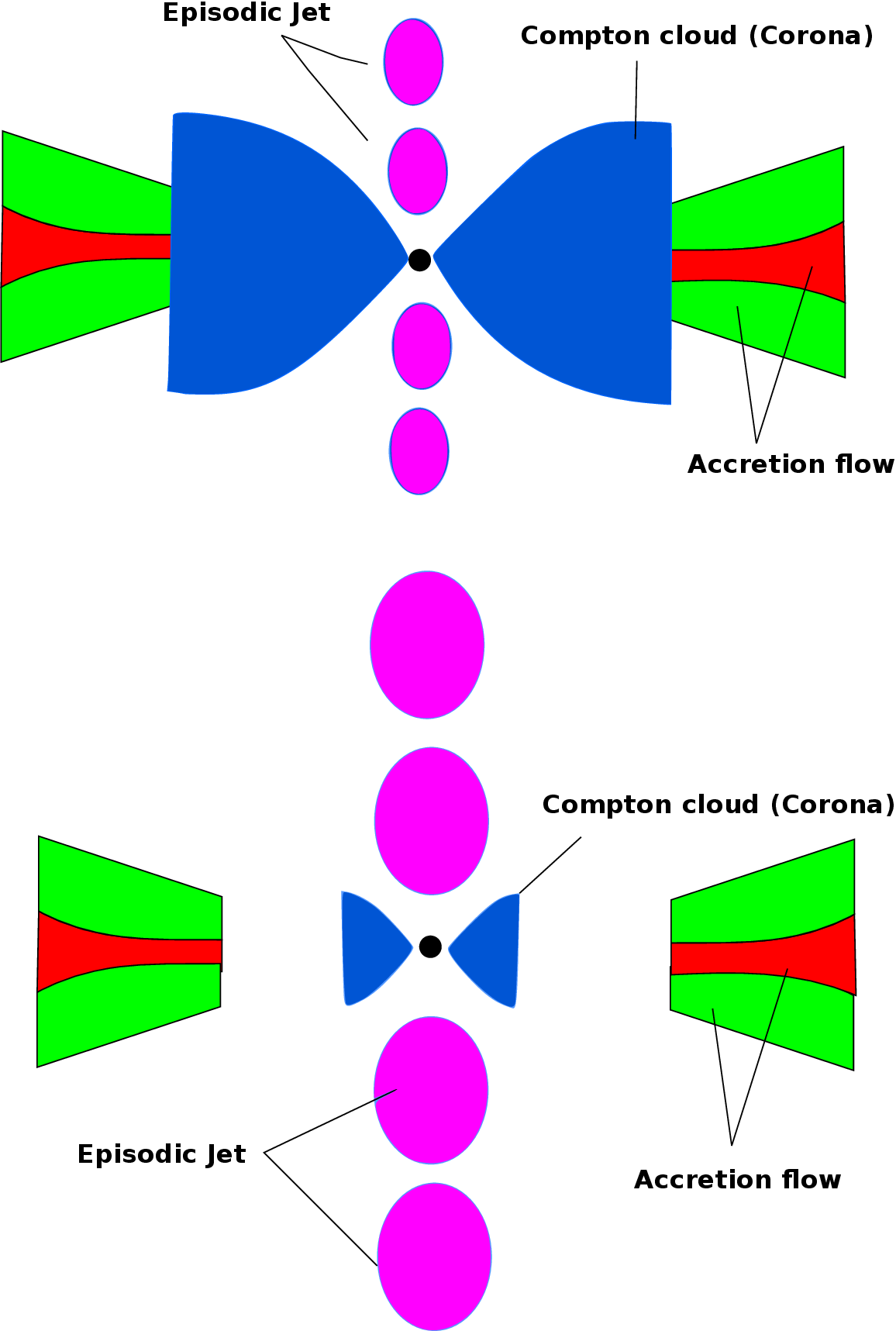}
    \caption{Cartoon diagram representing the accretion geometry in GX~339--4. A more extensive corona with episodic jet ejection is shown in the top panel, whereas the bottom panel shows a shrunk corona due to increased mass ejection rate. In both panels, the corona is represented in blue shaded region, whereas the magenta-coloured blobs represent the episodic jet. The sub-Keplerian and Keplerian accretion flows are represented with red- and green-coloured regions, respectively.}
    \label{fig:cartoon}
\end{figure}

\subsection{Evolution of the Inner Accretion Flow}
\label{sec:inner-flow}

At the beginning of the outburst, the LHS is observed, with the accretion disc truncated at a larger distance. In this state, the Comptonized emission dominates the spectrum with $f_{\rm disc}<20\%$ \citep[e.g.,][]{RM06}. As the outburst progresses, the disc emission starts to dominate, with the disc moving inward and the size of the Compton corona shrinking. The source moves to the HSS via IMS. The source enters the declining phase when the viscosity is turned off. As a result, the matter supply is cut-off, and the source moves to the quiescent state via IMS and LHS. In the IMS, a blobby jet or discrete ejection is observed. The Compton corona acts as the base of the jet. As a blob moves away at a relativistic speed, suddenly, a void is created at the Compton corona. The disc moves quickly to fill the void making the spectrum soft for a short time. As the corona recovers, the spectrum becomes harder. As a result, we observed fluctuations in the X-ray flux, $T_{\rm in}$, and $\Gamma$ in a very short timescale in the IMS.

At the beginning of the \astrosat~ observation, GX~339--4 was in the SIMS, with $T_{\rm in} \sim 1$~keV, and $\Gamma \sim 2.3$. The source moved to the HIMS at the end of the observation period. The source showed an AFP around MJD 59305–59306 as the X-ray flux decreased. The disc slowly moved outward as $T_{\rm in}$ decreased to $\sim 0.7$~keV. The spectrum became hard with decreasing $\Gamma$. In this case, one would expect $f_{\rm disc}$ to decrease. However, we observed an opposite scenario, where $f_{\rm disc}$ increases with a decrease in $\Gamma$. This is observed as $F_{\rm PL}$ is reduced, although $F_{\rm DBB}$ remains the same. During the observation period, the $\dot{m}_{\rm d}$ was constant within uncertainty, which is responsible for a constant $F_{\rm DBB}$. 

The AFP can be explained in terms of an episodic ejection, which is observed in the IMS. If a blobby jet is produced, a part of the matter from the corona is ejected, leaving the size of the corona shrinking, which in turn, reduces $F_{\rm PL}$, resulting an AFP. At the same time, as the size of the corona decreases, the number of intercepted soft photons decreases, leading to inefficient cooling, which leads to the harder spectrum, i.e., decreasing $\Gamma$. Similar behavior is also observed in other BHXBs \citep[][and references therein]{Vadawale2001}. 

Above mentioned scenario is observed in GX~339--4. Initially, the coronal size was $\sim 75$ $R_g$ with mass ejection rate as $\sim 0.2$ $\dot{m}_{\rm Edd}$ on MJD 59303. The jet flux was estimated to increase on MJD 59305 with $\dot{m}_{\rm out} \sim 0.33$ $\dot{m}_{\rm Edd}$, indicating a huge amount of the mass ejection on that day. As a result, the corona size decreased to $\sim 50$ $R_g$ on the next day, on MJD 59306. After that, the corona remains at $\sim 50$~$R_g$, with the mass ejection rate increased to $\sim 0.5$ $\dot{m}_{\rm out}$ at the end of our observation period. As more mass is ejected, the corona becomes weak with decreasing Comptonized flux. Figure~\ref{fig:cartoon} shows the cartoon diagram of the inner accretion flow in GX~339--4. The top panel shows a larger corona with episodic jet ejection. With the increase in the mass ejection rate, the mass from the corona is ejected, leaving behind a shrunken corona.

\citet{Liu2022} studied the source on MJD 59303--59305 with Insight-HXMT, and suggested that the AFP is seen due to the change in the coronal power in a short timescale. The authors suggested that there was a sudden change in the magnetic field, which led to a sudden heating effect in the corona, which reduced the coronal power. The authors did not find any variation in the coronal geometry, which contradicts our findings. The authors assumed a constant inner disc radius throughout their observation and found no variation of the coronal height \citep[see, Table~2 of][]{Liu2022}. However, in our analysis, we allowed the inner disc radius to vary and found both $R_{\rm in}$ and $h$ vary, indicating a dynamic inner accretion region.

Although, \citet{Peirano2023} did not study the AFP period, they suggested the existence of dual corona: a small compact corona ($\sim 25~R_g$) with high feedback and an extended corona ($\sim 1500~R_g$) with no feedback. \citet{Yang2023} argued that the extended corona could be the base of the jet and can explain the variation of type-B QPO. However, the QPOs can also be explained by a single corona model \citep{MondalEtal2023arXiv230303742M}. If the dual corona existed in GX~339--4, the outer corona could be ejected, leaving a smaller compact inner corona, which also supports our explanation of AFP.

\subsection{Reprocessed Emission}
The reprocessed emission produced a Compton hump at $\sim 20-40$~keV and a Fe~K$\alpha$ line at 6.4 keV. As we carried out our spectral analysis up to 25 keV, the Compton hump was not observed in the spectra. However, the Fe~K$\alpha$ line was detected at 6.4 keV while fitting the orbit-wise spectra with Model-1. We used the EW of the Fe~K$\alpha$ line to study the reprocessed emission in the source. Figure~\ref{fig:eq} shows the variation of the EW as a function of $\Gamma$. We observed that the EW is correlated with $\Gamma$. This indicates that the reflection becomes stronger when the disc moves inward. This enables more hard photons to intercept the disc, resulting in increasing reprocessed emission, thereby, the EW. However, in the HSS, the coronal emission decreases, leading to a decrease in the reprocessed emission, i.e., EW. Generally, the EW is observed to correlate with the photon index up to $\Gamma \sim 2$ while the reprocessed emission saturates at $\Gamma >2$; and decreases with further increasing $\Gamma$ \citep[e.g.,][and references therein]{AJ2022c}. A similar variation of the reflection is also observed in several active galactic nuclei and other BHXBs \citep{Zdziarski1999,Gilfanov1999,Lubinski2001,Ezhikode2020}.

We used the relativistic reflection model \textsc{relxillLp} to analyze one-day-averaged spectra. The reflection fraction ($R_f$) is not a free parameter in \textsc{relxillLp}, as it is estimated self-consistently. Hence, we estimated the flux of the reprocessed emission ($F_{\rm ref}$) and primary emission ($F_{\rm PL}$) in $10-25$~keV to compute the reflection fraction ($R_{\rm f}=F_{\rm ref}/F_{\rm PL}$). We found that the $R_f$ decreases as the outburst progressed, following the similar variation of the EW of iron K$\alpha$ line. The reprocessed emission generally depends on the disc ionization and hard X-ray emission. As disk ionization was constant during the observation period at $\log \xi \sim 10^4$ \ecps, the hard coronal X-ray is responsible for the change in the reprocessed emission. The decreasing $F_{\rm PL}$ naturally explains the decreasing the $R_{\rm {\rm f}}$, and EW.

\section{Summary}
We studied spectral properties of the black hole binary GX~339--4 using \astrosat~ observations in $0.7-25$~keV energy ranges during its 2021 outburst. Our key findings are given below.

\begin{enumerate}
\item The 2021 outburst of GX~339--4 lasted for about 10 months. \astrosat~ observed the source in the intermediate state during the rising phase of the outburst.
\item The mass of BH is estimated to be $7.3 \pm 0.6$ $M_{\odot}$.
\item An `alternate flux profile' was seen during our observation period, which can be explained with an episodic jet ejection.
\item The mass ejection rate was found to reach the peak when AFP was observed. As more mass is ejected, the corona becomes weak with decreasing Comptonized flux.
\item During our observation period, we observed a dynamic inner accretion flow, with a contracting corona, and a receding disc.
\item A broad iron k$\alpha$ line was observed in the spectrum. The EW of the iron line is found to correlate with the photon index $\Gamma$. The reflection fraction is found to decrease with the decrease in the Comptonized flux.
\item The disc did not follow the $F_{\rm DBB}-T_{\rm in}^4$ relation during the \astrosat~ observation of the source, which is an indication of a moving disc.
    
\end{enumerate}

\section*{Acknowledgements}
We thank the anonymous reviewer's for his/her comments and suggestions which helped to improve the paper. AJ and HK acknowledge the support of the grant from the Ministry of Science and Technology of Taiwan with the grant numbers MOST~110-2811-M-007-500 and MOST~111-2811-M-007-002. HK acknowledge the support of the grant from the Ministry of Science and Technology of Taiwan with the grand number MOST~110-2112-M-007-020 and MOST-111-2112-M-007-019. SM acknowledges Ramanujan Fellowship research grant (\#RJF/2020/000113) by SERB-DST, Government of India for this work. Work at Physical Research Laboratory, Ahmedabad, is funded by the Department of Space, Government of India. This research made use of the data obtained through ToO phase of AstroSat observations. The authors thank the SXT-POC of TIFR and the LAXPC team of IUCAA and TIFR for providing the data extraction software for the respective instruments.

\section*{DATA AVAILABILITY}
We used the data of \astrosat~ observatories for this work. 
\bibliographystyle{mnras}
\bibliography{ref-gx339}


\appendix
\section{Table-A}

\begin{table*}
\caption{Best-fit parameters obtained from the spectral fitting of data obtained from the combined SXT and LAXPC observations in the $0.7-25$~keV energy range.}
\label{tab:spec-res}
\centering
\begin{tabular}{c|c|c|c|c|c|c|c|c|c|}
\hline
Orbit& Avg. MJD & $N_{\rm H}$ & $T_{\rm in}$ & $R_{\rm in}$  & $\Gamma$ & EW & $\chi^2$/dof & $F_{\rm tot}$ & $f_{\rm disc}$\\
    &            & ($10^{22}$ \pcm) & (keV) &   (km) &       & (eV) &  &($10^{-8}$ \ecps ) &          \\    
\hline
29750& 59303.06&$ 0.78^{+0.02}_{-0.02}$&$ 0.96^{+0.02}_{-0.01}$& $  33.34^{+1.43}_{-1.60}   $ & $ 1.99^{+0.04}_{-0.03}$&$  925^{+28}_{-33}$& 617/639&$  2.12\pm 0.02$&$ 0.50\pm 0.01 $\\ 
29755& 59303.12&$ 0.75^{+0.02}_{-0.03}$&$ 0.97^{+0.02}_{-0.02}$& $  30.31^{+1.48}_{-1.64}   $ & $ 2.00^{+0.02}_{-0.05}$&$  926^{+25}_{-35}$& 617/639&$  2.22\pm 0.02$&$ 0.49\pm 0.01 $\\
29756& 59303.34&$ 0.78^{+0.03}_{-0.02}$&$ 0.93^{+0.02}_{-0.02}$& $  32.46^{+1.22}_{-1.38}   $ & $ 2.02^{+0.02}_{-0.04}$&$  925^{+20}_{-26}$& 569/636&$  2.01\pm 0.02$&$ 0.52\pm 0.01 $\\
29758& 59303.54&$ 0.74^{+0.03}_{-0.01}$&$ 0.97^{+0.01}_{-0.02}$& $  31.70^{+2.06}_{-1.78}   $ & $ 2.04^{+0.04}_{-0.03}$&$  934^{+17}_{-21}$& 594/639&$  2.14\pm 0.02$&$ 0.50\pm 0.01 $\\
29759& 59303.61&$ 0.79^{+0.02}_{-0.02}$&$ 0.95^{+0.02}_{-0.01}$& $  33.53^{+2.24}_{-2.50}   $ & $ 2.02^{+0.03}_{-0.02}$&$  876^{+21}_{-26}$& 589/639&$  2.10\pm 0.02$&$ 0.50\pm 0.01 $\\
29760& 59303.69&$ 0.77^{+0.01}_{-0.02}$&$ 0.93^{+0.03}_{-0.02}$& $  34.25^{+1.99}_{-1.48}   $ & $ 2.00^{+0.03}_{-0.02}$&$  875^{+16}_{-19}$& 690/638&$  2.04\pm 0.02$&$ 0.51\pm 0.02 $\\
29761& 59303.77&$ 0.79^{+0.02}_{-0.01}$&$ 0.93^{+0.02}_{-0.02}$& $  36.86^{+1.69}_{-1.24}   $ & $ 1.98^{+0.02}_{-0.04}$&$  843^{+19}_{-25}$& 607/638&$  2.00\pm 0.02$&$ 0.51\pm 0.02 $\\
29763& 59303.91&$ 0.80^{+0.01}_{-0.01}$&$ 0.94^{+0.02}_{-0.03}$& $  34.23^{+1.65}_{-1.98}   $ & $ 2.00^{+0.04}_{-0.03}$&$  871^{+22}_{-33}$& 657/639&$  2.04\pm 0.02$&$ 0.51\pm 0.01 $\\
29764& 59303.49&$ 0.78^{+0.02}_{-0.01}$&$ 0.94^{+0.02}_{-0.02}$& $  32.17^{+2.15}_{-2.33}   $ & $ 1.98^{+0.02}_{-0.03}$&$  890^{+18}_{-17}$& 686/639&$  2.01\pm 0.02$&$ 0.52\pm 0.02 $\\
29765& 59304.05&$ 0.78^{+0.01}_{-0.02}$&$ 1.00^{+0.02}_{-0.02}$& $  29.34^{+1.40}_{-1.62}   $ & $ 1.97^{+0.02}_{-0.02}$&$  902^{+20}_{-24}$& 661/638&$  1.99\pm 0.01$&$ 0.57\pm 0.01 $\\
29770& 59304.20&$ 0.74^{+0.02}_{-0.01}$&$ 1.00^{+0.03}_{-0.02}$& $  32.40^{+1.33}_{-1.71}   $ & $ 1.95^{+0.03}_{-0.02}$&$  924^{+23}_{-28}$& 665/635&$  2.04\pm 0.02$&$ 0.55\pm 0.02 $\\
29776& 59304.78&$ 0.78^{+0.02}_{-0.03}$&$ 0.95^{+0.02}_{-0.03}$& $  33.79^{+1.30}_{-1.66}   $ & $ 1.94^{+0.02}_{-0.03}$&$  924^{+28}_{-35}$& 604/639&$  1.94\pm 0.02$&$ 0.57\pm 0.02 $\\ 
29777& 59304.85&$ 0.78^{+0.01}_{-0.02}$&$ 0.99^{+0.02}_{-0.02}$& $  32.44^{+1.56}_{-1.96}   $ & $ 1.94^{+0.03}_{-0.02}$&$  865^{+43}_{-49}$& 679/637&$  1.98\pm 0.02$&$ 0.56\pm 0.02 $\\ 
29778& 59304.93&$ 0.78^{+0.02}_{-0.01}$&$ 0.98^{+0.03}_{-0.01}$& $  32.95^{+2.23}_{-2.48}   $ & $ 1.95^{+0.02}_{-0.02}$&$  864^{+35}_{-48}$& 621/638&$  2.03\pm 0.02$&$ 0.56\pm 0.02 $\\ 
29779& 59304.50&$ 0.78^{+0.01}_{-0.02}$&$ 0.98^{+0.03}_{-0.02}$& $  35.39^{+1.92}_{-1.37}   $ & $ 1.94^{+0.01}_{-0.03}$&$  902^{+28}_{-41}$& 672/638&$  2.03\pm 0.02$&$ 0.55\pm 0.02 $\\ 
29784& 59305.05&$ 0.78^{+0.01}_{-0.03}$&$ 0.97^{+0.02}_{-0.01}$& $  37.14^{+2.03}_{-2.29}   $ & $ 1.93^{+0.02}_{-0.04}$&$  925^{+26}_{-36}$& 661/638&$  2.00\pm 0.01$&$ 0.55\pm 0.01 $\\ 
29785& 59305.27&$ 0.77^{+0.02}_{-0.01}$&$ 0.97^{+0.02}_{-0.03}$& $  37.75^{+2.13}_{-1.69}   $ & $ 1.92^{+0.04}_{-0.03}$&$  920^{+45}_{-52}$& 662/639&$  1.94\pm 0.02$&$ 0.57\pm 0.02 $\\ 
29787& 59305.28&$ 0.73^{+0.02}_{-0.02}$&$ 0.96^{+0.02}_{-0.02}$& $  41.58^{+2.00}_{-2.16}   $ & $ 1.90^{+0.04}_{-0.05}$&$  653^{+23}_{-28}$& 698/639&$  1.86\pm 0.02$&$ 0.62\pm 0.02 $\\ 
29788& 59305.57&$ 0.72^{+0.02}_{-0.01}$&$ 0.91^{+0.03}_{-0.03}$& $  42.98^{+1.50}_{-1.83}   $ & $ 1.89^{+0.03}_{-0.03}$&$  842^{+19}_{-25}$& 664/638&$  1.63\pm 0.01$&$ 0.60\pm 0.02 $\\ 
29789& 59305.66&$ 0.74^{+0.01}_{-0.02}$&$ 0.95^{+0.03}_{-0.02}$& $  39.88^{+1.37}_{-1.67}   $ & $ 1.92^{+0.03}_{-0.04}$&$  824^{+42}_{-53}$& 625/638&$  1.99\pm 0.03$&$ 0.57\pm 0.02 $\\ 
29790& 59305.73&$ 0.76^{+0.02}_{-0.03}$&$ 0.86^{+0.02}_{-0.02}$& $  42.08^{+2.50}_{-2.03}   $ & $ 1.95^{+0.04}_{-0.03}$&$  833^{+37}_{-45}$& 628/639&$  1.59\pm 0.02$&$ 0.61\pm 0.03 $\\ 
29791& 59305.80&$ 0.78^{+0.01}_{-0.01}$&$ 0.85^{+0.03}_{-0.01}$& $  42.00^{+1.59}_{-1.69}   $ & $ 1.95^{+0.02}_{-0.03}$&$  899^{+41}_{-48}$& 674/639&$  1.61\pm 0.01$&$ 0.60\pm 0.01 $\\ 
29792& 59305.86&$ 0.76^{+0.01}_{-0.02}$&$ 0.86^{+0.02}_{-0.02}$& $  42.03^{+1.66}_{-2.07}   $ & $ 1.96^{+0.03}_{-0.04}$&$  880^{+33}_{-39}$& 639/639&$  1.64\pm 0.01$&$ 0.60\pm 0.02 $\\ 
29793& 59305.88&$ 0.78^{+0.01}_{-0.02}$&$ 0.86^{+0.02}_{-0.02}$& $  44.17^{+1.91}_{-2.19}   $ & $ 1.94^{+0.03}_{-0.04}$&$  867^{+47}_{-58}$& 667/639&$  1.67\pm 0.02$&$ 0.61\pm 0.02 $\\ 
29794& 59305.51&$ 0.76^{+0.02}_{-0.02}$&$ 0.92^{+0.02}_{-0.01}$& $  40.46^{+1.52}_{-1.67}   $ & $ 1.94^{+0.03}_{-0.04}$&$  893^{+41}_{-46}$& 624/639&$  1.72\pm 0.03$&$ 0.61\pm 0.02 $\\ 
29799& 59306.13&$ 0.75^{+0.03}_{-0.03}$&$ 0.89^{+0.01}_{-0.02}$& $  40.62^{+2.52}_{-2.67}   $ & $ 1.94^{+0.03}_{-0.03}$&$  860^{+18}_{-22}$& 648/639&$  1.67\pm 0.02$&$ 0.62\pm 0.03 $\\ 
29801& 59306.34&$ 0.76^{+0.01}_{-0.02}$&$ 0.88^{+0.03}_{-0.01}$& $  41.93^{+1.37}_{-1.58}   $ & $ 1.89^{+0.04}_{-0.05}$&$  791^{+25}_{-29}$& 657/633&$  1.52\pm 0.03$&$ 0.63\pm 0.03 $\\ 
29802& 59306.51&$ 0.79^{+0.01}_{-0.01}$&$ 0.87^{+0.02}_{-0.02}$& $  43.73^{+1.95}_{-1.58}   $ & $ 1.88^{+0.03}_{-0.02}$&$  758^{+28}_{-37}$& 673/635&$  1.25\pm 0.02$&$ 0.51\pm 0.02 $\\ 
29803& 59306.58&$ 0.73^{+0.02}_{-0.03}$&$ 0.92^{+0.03}_{-0.02}$& $  41.48^{+2.11}_{-1.73}   $ & $ 1.91^{+0.04}_{-0.03}$&$  866^{+24}_{-30}$& 701/639&$  1.61\pm 0.02$&$ 0.61\pm 0.02 $\\ 
29804& 59306.65&$ 0.78^{+0.01}_{-0.01}$&$ 0.90^{+0.02}_{-0.01}$& $  41.29^{+1.47}_{-1.67}   $ & $ 1.91^{+0.02}_{-0.03}$&$  938^{+18}_{-24}$& 622/638&$  1.56\pm 0.01$&$ 0.63\pm 0.01 $\\ 
29805& 59306.74&$ 0.76^{+0.03}_{-0.03}$&$ 0.95^{+0.02}_{-0.02}$& $  39.14^{+1.61}_{-1.90}   $ & $ 1.95^{+0.04}_{-0.03}$&$  886^{+21}_{-25}$& 626/639&$  1.92\pm 0.02$&$ 0.59\pm 0.01 $\\ 
29806& 59306.82&$ 0.78^{+0.02}_{-0.02}$&$ 0.96^{+0.01}_{-0.02}$& $  36.65^{+2.39}_{-2.69}   $ & $ 1.97^{+0.04}_{-0.02}$&$  928^{+27}_{-33}$& 622/638&$  1.98\pm 0.02$&$ 0.58\pm 0.01 $\\ 
29807& 59306.89&$ 0.78^{+0.01}_{-0.02}$&$ 0.97^{+0.01}_{-0.01}$& $  35.45^{+1.98}_{-2.24}   $ & $ 1.96^{+0.03}_{-0.03}$&$  827^{+33}_{-36}$& 708/638&$  2.02\pm 0.02$&$ 0.58\pm 0.02 $\\ 
29808& 59306.94&$ 0.78^{+0.02}_{-0.02}$&$ 0.99^{+0.02}_{-0.02}$& $  33.60^{+2.35}_{-2.69}   $ & $ 1.97^{+0.02}_{-0.03}$&$  801^{+39}_{-48}$& 619/639&$  2.06\pm 0.02$&$ 0.58\pm 0.02 $\\ 
29814& 59306.61&$ 0.77^{+0.02}_{-0.02}$&$ 0.97^{+0.01}_{-0.02}$& $  32.15^{+2.10}_{-2.20}   $ & $ 1.96^{+0.03}_{-0.04}$&$  896^{+15}_{-19}$& 681/639&$  2.09\pm 0.02$&$ 0.57\pm 0.02 $\\ 
29816& 59307.35&$ 0.78^{+0.01}_{-0.03}$&$ 0.91^{+0.02}_{-0.01}$& $  38.00^{+2.08}_{-1.57}   $ & $ 1.95^{+0.02}_{-0.04}$&$  834^{+25}_{-29}$& 626/639&$  1.91\pm 0.02$&$ 0.59\pm 0.02 $\\ 
29817& 59307.53&$ 0.75^{+0.03}_{-0.03}$&$ 0.94^{+0.02}_{-0.02}$& $  36.41^{+2.34}_{-1.75}   $ & $ 1.97^{+0.03}_{-0.03}$&$  836^{+23}_{-30}$& 629/646&$  2.00\pm 0.02$&$ 0.59\pm 0.02 $\\ 
29818& 59307.60&$ 0.77^{+0.02}_{-0.03}$&$ 0.93^{+0.02}_{-0.02}$& $  40.15^{+1.87}_{-2.33}   $ & $ 1.98^{+0.03}_{-0.03}$&$  881^{+22}_{-30}$& 626/639&$  2.10\pm 0.02$&$ 0.62\pm 0.01 $\\ 
29819& 59307.69&$ 0.79^{+0.01}_{-0.03}$&$ 0.89^{+0.02}_{-0.01}$& $  41.58^{+1.80}_{-2.12}   $ & $ 1.94^{+0.04}_{-0.03}$&$  825^{+32}_{-26}$& 612/646&$  1.78\pm 0.02$&$ 0.57\pm 0.02 $\\ 
29820& 59307.76&$ 0.79^{+0.02}_{-0.02}$&$ 0.76^{+0.02}_{-0.02}$& $  43.84^{+1.93}_{-2.11}   $ & $ 1.92^{+0.04}_{-0.02}$&$  747^{+38}_{-46}$& 681/638&$  1.70\pm 0.01$&$ 0.60\pm 0.01 $\\ 
29821& 59307.83&$ 0.78^{+0.02}_{-0.02}$&$ 0.77^{+0.03}_{-0.02}$& $  43.31^{+1.50}_{-1.59}   $ & $ 1.91^{+0.02}_{-0.03}$&$  718^{+40}_{-30}$& 700/638&$  1.66\pm 0.01$&$ 0.62\pm 0.02 $\\ 
29822& 59307.90&$ 0.79^{+0.02}_{-0.01}$&$ 0.73^{+0.03}_{-0.02}$& $  41.87^{+1.55}_{-1.75}   $ & $ 1.87^{+0.03}_{-0.02}$&$  790^{+39}_{-32}$& 693/646&$  1.67\pm 0.02$&$ 0.63\pm 0.02 $\\ 
29823& 59307.47&$ 0.76^{+0.03}_{-0.02}$&$ 0.74^{+0.02}_{-0.03}$& $  41.56^{+1.82}_{-1.84}   $ & $ 1.97^{+0.03}_{-0.02}$&$  794^{+38}_{-45}$& 691/643&$  1.73\pm 0.02$&$ 0.61\pm 0.02 $\\ 
29828& 59307.81&$ 0.74^{+0.03}_{-0.02}$&$ 0.79^{+0.01}_{-0.03}$& $  40.95^{+2.20}_{-2.47}   $ & $ 1.92^{+0.04}_{-0.03}$&$  681^{+45}_{-57}$& 696/639&$  1.65\pm 0.03$&$ 0.64\pm 0.03 $\\ 
29830& 59308.27&$ 0.78^{+0.02}_{-0.02}$&$ 0.78^{+0.03}_{-0.02}$& $  41.94^{+2.43}_{-1.88}   $ & $ 1.93^{+0.02}_{-0.03}$&$  771^{+29}_{-31}$& 670/639&$  1.72\pm 0.03$&$ 0.62\pm 0.03 $\\ 
\hline
\end{tabular}
\end{table*}

\begin{table*}
\contcaption{Best-fit parameters obtained from the spectral fitting of data obtained from the combined SXT and LAXPC observations in the $0.7-25$~keV energy range.}
\centering
\begin{tabular}{c|c|c|c|c|c|c|c|c|c|}
\hline
Orbit& Avg. MJD & $N_{\rm H}$ & $T_{\rm in}$ & $R_{\rm in}$  & $\Gamma$ & EW & $\chi^2$/dof & $F_{\rm tot}$ & $f_{\rm disc}$\\
    &            & ($10^{22}$ \pcm) & (keV) &   (km) &       & (eV) &  &($10^{-8}$ \ecps ) &          \\    
\hline
29831& 59308.47&$ 0.77^{+0.01}_{-0.02}$&$ 0.78^{+0.02}_{-0.03}$& $  41.44^{+2.60}_{-2.77}   $ & $ 1.93^{+0.03}_{-0.02}$&$  847^{+30}_{-25}$& 679/638&$  1.73\pm 0.02$&$ 0.61\pm 0.03 $\\ 
29832& 59308.54&$ 0.74^{+0.03}_{-0.01}$&$ 0.77^{+0.02}_{-0.02}$& $  43.29^{+2.83}_{-2.25}   $ & $ 1.94^{+0.03}_{-0.02}$&$  842^{+24}_{-31}$& 668/639&$  1.76\pm 0.02$&$ 0.61\pm 0.03 $\\ 
29833& 59308.61&$ 0.73^{+0.02}_{-0.02}$&$ 0.75^{+0.03}_{-0.02}$& $  43.43^{+1.67}_{-2.07}   $ & $ 1.94^{+0.04}_{-0.02}$&$  903^{+35}_{-27}$& 647/641&$  1.71\pm 0.02$&$ 0.61\pm 0.03 $\\ 
29834& 59308.70&$ 0.78^{+0.01}_{-0.03}$&$ 0.73^{+0.02}_{-0.01}$& $  45.26^{+1.65}_{-2.06}   $ & $ 1.93^{+0.03}_{-0.04}$&$  722^{+26}_{-35}$& 674/638&$  1.67\pm 0.02$&$ 0.61\pm 0.02 $\\ 
29835& 59308.77&$ 0.79^{+0.02}_{-0.02}$&$ 0.73^{+0.03}_{-0.02}$& $  45.15^{+2.03}_{-2.36}   $ & $ 1.93^{+0.04}_{-0.02}$&$  745^{+29}_{-41}$& 631/639&$  1.71\pm 0.02$&$ 0.62\pm 0.02 $\\ 
29836& 59308.85&$ 0.76^{+0.02}_{-0.02}$&$ 0.72^{+0.02}_{-0.01}$& $  44.96^{+2.18}_{-2.46}   $ & $ 1.89^{+0.03}_{-0.03}$&$  716^{+31}_{-38}$& 667/643&$  1.64\pm 0.03$&$ 0.61\pm 0.03 $\\ 
29837& 59308.92&$ 0.77^{+0.02}_{-0.02}$&$ 0.70^{+0.02}_{-0.01}$& $  47.90^{+2.23}_{-2.44}   $ & $ 1.88^{+0.04}_{-0.02}$&$  720^{+25}_{-30}$& 686/645&$  1.53\pm 0.02$&$ 0.62\pm 0.02 $\\ 
29838& 59308.49&$ 0.80^{+0.02}_{-0.02}$&$ 0.71^{+0.01}_{-0.03}$& $  47.66^{+2.14}_{-2.51}   $ & $ 1.88^{+0.02}_{-0.03}$&$  711^{+40}_{-32}$& 685/649&$  1.63\pm 0.03$&$ 0.65\pm 0.04 $\\ 
29843& 59309.12&$ 0.81^{+0.02}_{-0.03}$&$ 0.72^{+0.01}_{-0.02}$& $  42.68^{+1.58}_{-1.93}   $ & $ 1.85^{+0.04}_{-0.04}$&$  708^{+36}_{-46}$& 622/647&$  1.43\pm 0.03$&$ 0.58\pm 0.04 $\\ 
29845& 59309.34&$ 0.79^{+0.03}_{-0.02}$&$ 0.73^{+0.02}_{-0.01}$& $  45.99^{+1.65}_{-2.05}   $ & $ 1.86^{+0.03}_{-0.02}$&$  690^{+28}_{-32}$& 711/638&$  1.65\pm 0.02$&$ 0.66\pm 0.04 $\\ 
29846& 59309.49&$ 0.73^{+0.03}_{-0.03}$&$ 0.74^{+0.03}_{-0.02}$& $  42.71^{+1.86}_{-2.31}   $ & $ 1.86^{+0.04}_{-0.03}$&$  654^{+18}_{-23}$& 663/649&$  1.66\pm 0.02$&$ 0.67\pm 0.03 $\\ 
29847& 59309.56&$ 0.75^{+0.02}_{-0.02}$&$ 0.73^{+0.02}_{-0.01}$& $  45.15^{+1.80}_{-2.33}   $ & $ 1.84^{+0.04}_{-0.02}$&$  607^{+19}_{-28}$& 634/645&$  1.63\pm 0.02$&$ 0.68\pm 0.02 $\\ 
29848& 59309.63&$ 0.80^{+0.02}_{-0.02}$&$ 0.74^{+0.01}_{-0.02}$& $  43.43^{+1.43}_{-1.86}   $ & $ 1.86^{+0.02}_{-0.03}$&$  631^{+26}_{-36}$& 692/646&$  1.64\pm 0.02$&$ 0.67\pm 0.02 $\\ 
29849& 59309.72&$ 0.74^{+0.03}_{-0.01}$&$ 0.73^{+0.02}_{-0.01}$& $  44.85^{+1.48}_{-1.84}   $ & $ 1.87^{+0.02}_{-0.02}$&$  611^{+21}_{-29}$& 611/648&$  1.57\pm 0.04$&$ 0.65\pm 0.04 $\\ 
29850& 59309.78&$ 0.81^{+0.02}_{-0.02}$&$ 0.74^{+0.02}_{-0.01}$& $  44.62^{+1.35}_{-1.55}   $ & $ 1.87^{+0.03}_{-0.04}$&$  626^{+26}_{-21}$& 612/642&$  1.67\pm 0.03$&$ 0.67\pm 0.03 $\\ 
29851& 59309.86&$ 0.76^{+0.03}_{-0.03}$&$ 0.75^{+0.03}_{-0.01}$& $  44.40^{+1.35}_{-1.16}   $ & $ 1.88^{+0.03}_{-0.02}$&$  637^{+28}_{-35}$& 669/631&$  1.67\pm 0.02$&$ 0.67\pm 0.02 $\\ 
29852& 59309.94&$ 0.78^{+0.02}_{-0.01}$&$ 0.74^{+0.02}_{-0.02}$& $  44.09^{+1.56}_{-1.69}   $ & $ 1.85^{+0.02}_{-0.03}$&$  609^{+23}_{-26}$& 612/643&$  1.68\pm 0.02$&$ 0.66\pm 0.03 $\\ 
29857& 59309.48&$ 0.78^{+0.02}_{-0.01}$&$ 0.72^{+0.02}_{-0.01}$& $  46.02^{+1.68}_{-1.46}   $ & $ 1.84^{+0.03}_{-0.04}$&$  540^{+28}_{-39}$& 623/647&$  1.62\pm 0.02$&$ 0.67\pm 0.03 $\\ 
29859& 59309.63&$ 0.80^{+0.02}_{-0.02}$&$ 0.75^{+0.01}_{-0.02}$& $  43.92^{+1.84}_{-2.09}   $ & $ 1.84^{+0.02}_{-0.02}$&$  626^{+17}_{-23}$& 674/645&$  1.65\pm 0.02$&$ 0.67\pm 0.02 $\\ 
29860& 59310.39&$ 0.81^{+0.01}_{-0.02}$&$ 0.75^{+0.01}_{-0.02}$& $  43.89^{+1.34}_{-1.49}   $ & $ 1.87^{+0.02}_{-0.03}$&$  616^{+20}_{-28}$& 634/643&$  1.66\pm 0.02$&$ 0.67\pm 0.03 $\\ 
29861& 59310.50&$ 0.78^{+0.02}_{-0.02}$&$ 0.75^{+0.02}_{-0.01}$& $  43.77^{+1.28}_{-1.53}   $ & $ 1.84^{+0.04}_{-0.03}$&$  581^{+22}_{-25}$& 677/648&$  1.45\pm 0.03$&$ 0.63\pm 0.03 $\\ 
29862& 59310.57&$ 0.74^{+0.03}_{-0.02}$&$ 0.76^{+0.02}_{-0.03}$& $  43.63^{+1.64}_{-1.87}   $ & $ 1.83^{+0.02}_{-0.02}$&$  603^{+30}_{-37}$& 691/639&$  1.65\pm 0.02$&$ 0.67\pm 0.02 $\\ 
29863& 59310.66&$ 0.75^{+0.03}_{-0.03}$&$ 0.74^{+0.01}_{-0.02}$& $  43.25^{+1.97}_{-2.14}   $ & $ 1.83^{+0.02}_{-0.03}$&$  582^{+29}_{-36}$& 688/642&$  1.63\pm 0.02$&$ 0.68\pm 0.03 $\\
29864& 59310.73&$ 0.78^{+0.02}_{-0.01}$&$ 0.75^{+0.02}_{-0.01}$& $  42.58^{+1.44}_{-1.56}   $ & $ 1.82^{+0.03}_{-0.04}$&$  632^{+25}_{-32}$& 672/639&$  1.58\pm 0.02$&$ 0.68\pm 0.03 $\\
29865& 59310.81&$ 0.79^{+0.02}_{-0.02}$&$ 0.74^{+0.01}_{-0.01}$& $  42.27^{+1.52}_{-1.70}   $ & $ 1.80^{+0.02}_{-0.03}$&$  543^{+27}_{-31}$& 671/649&$  1.58\pm 0.03$&$ 0.69\pm 0.04 $\\ 
\hline
\end{tabular}
\leftline{Errors are quoted at 90\% confidence.}
\end{table*}

\section{Spectra analysis result of one day-averaged spectra}

\begin{table*}
\caption{Best-fit parameters obtained from the spectral fitting of one day-averaged data with \textsc{relxillLp} model.}
\label{tab:relxill}
\centering
\begin{tabular}{c|c|c|c|c|c|c|c|c|c|c}
\hline
Orbit& MJD &  $T_{\rm in}$ & $\Gamma$ &  $R_{\rm in}$  & $h$ & $\log \xi$   & $A_{\rm Fe}$ & $R_{\rm f}$   &$\chi^2$/dof \\
    &            & (keV) &    & ($R_{\rm ISCO}$)   & ($R_{\rm g}$) &   (\ecps) & ($A_{\odot}$) &     \\    
    \hline
29750--29764& 59303&$ 0.95\pm0.02$&  $ 2.28\pm0.05$&$  1.7\pm0.1$& $  6.5^{+1.5}_{-0.8}$&$  3.9\pm0.4$& $  3.9\pm0.3$&$1.27\pm0.12$&767/632&\\ 
29765--29778& 59304&$ 0.98\pm0.02$&  $ 2.18\pm0.04$&$  1.5\pm0.1$& $  6.1^{+2.2}_{-1.1}$&$  4.1\pm0.4$& $  4.1\pm0.3$&$1.13\pm0.15$&687/632&\\
29779--29793& 59305&$ 0.92\pm0.02$&  $ 2.16\pm0.03$&$  2.5\pm0.2$& $  4.5^{+0.8}_{-0.7}$&$  3.9\pm0.3$& $  4.1\pm0.3$&$1.04\pm0.11$&728/632&\\
29794--29808& 59306&$ 0.91\pm0.02$&  $ 2.15\pm0.04$&$  2.1\pm0.1$& $  3.9^{+1.0}_{-0.7}$&$  4.1\pm0.2$& $  3.8\pm0.4$&$1.01\pm0.14$&683/632&\\
29809--29823& 59307&$ 0.94\pm0.02$&  $ 2.19\pm0.04$&$  2.8\pm0.2$& $  5.9^{+1.0}_{-0.8}$&$  3.9\pm0.3$& $  4.0\pm0.3$&$1.08\pm0.17$&708/632&\\
29838--29837& 59308&$ 0.75\pm0.03$&  $ 2.06\pm0.04$&$  3.9\pm0.2$& $  4.2^{+0.8}_{-0.7}$&$  4.0\pm0.3$& $  3.9\pm0.3$&$0.98\pm0.13$&710/632&\\
29839--29852& 59309&$ 0.73\pm0.02$&  $ 1.99\pm0.04$&$  4.3\pm0.2$& $  <2.7$&$  3.9\pm0.3$& $  4.2\pm0.3$&$0.91\pm0.12$&699/532&\\
29853--29865& 59310&$ 0.74\pm0.02$&  $ 1.97\pm0.04$&$  4.3\pm0.4$& $  <2.5      $&$  3.8\pm0.4$& $  4.0\pm0.3$&$0.91\pm0.14$&717/532&\\
\hline
\end{tabular}
\leftline{Errors are quoted at 90\% confidence (1.6$\sigma$).}
\end{table*}

\begin{table*}
\caption{Best-fit parameters obtained from the spectral fitting of one day-averaged data with \textsc{JeTCAF} model.}
\label{tab:tcaf}
\centering
\begin{tabular}{c|c|c|c|c|c|c|c|c|c|c}
\hline
Orbit& MJD &  $\dot{m}_{\rm d}$ & $\dot{m}_{\rm h}$ &  $X_{\rm S}$  & $R$ & $f_{\rm col}$   & $\dot{m}_{\rm out}$ &$\chi^2$/dof \\
    &        & ($\dot{m}_{\rm Edd}$)& ($\dot{m}_{\rm Edd}$)   & ($R_{\rm g}$)   & &  & ($\dot{m}_{\rm Edd}$) &         \\    
    \hline
29750--29764&59303&$1.87\pm  0.12$ &$0.21\pm 0.05$ &$75.6\pm  6.2  $&$3.50 \pm 0.44$ &$ 0.32\pm  0.05$ & $0.21\pm 0.02 $& 781/625 \\
29765--29778&59304&$2.07\pm  0.21$ &$0.24\pm 0.03$ &$57.9\pm  4.6  $&$2.96 \pm 0.31$ &$ 0.31\pm  0.06$ & $0.26\pm 0.06 $& 886/625 \\
29779--29793&59305&$1.77\pm  0.15$ &$0.21\pm 0.02$ &$57.3\pm  4.4  $&$3.29 \pm 0.21$ &$ 0.49\pm  0.10$ & $0.33\pm 0.07 $& 899/625 \\  
29794--29808&59306&$1.77\pm  0.14$ &$0.25\pm 0.06$ &$49.3\pm  5.3  $&$3.66 \pm 0.28$ &$ 0.41\pm  0.09$ & $0.25\pm 0.05 $& 898/625 \\
29809--29823&59307&$1.78\pm  0.13$ &$0.25\pm 0.06$ &$49.3\pm  4.9  $&$3.66 \pm 0.39$ &$ 0.39\pm  0.05$ & $0.24\pm 0.02 $& 895/625 \\
29838--29837&59308&$1.77\pm  0.08$ &$0.26\pm 0.04$ &$51.6\pm  3.9  $&$3.63 \pm 0.32$ &$ 0.60\pm  0.12$ & $0.37\pm 0.05 $& 897/625 \\
29839--29852&59309&$1.91\pm  0.17$ &$0.26\pm 0.06$ &$53.8\pm  5.7  $&$3.30 \pm 0.27$ &$ 0.68\pm  0.15$ & $0.49\pm 0.12 $& 864/625 \\
29853--29865&59310&$1.89\pm  0.19$ &$0.26\pm 0.05$ &$52.0\pm  4.5  $&$3.19 \pm 0.33$ &$ 0.73\pm  0.17$ & $0.54\pm 0.13 $& 810/625 \\
\hline
\end{tabular}
\leftline{Errors are quoted at 90\% confidence (1.6$\sigma$).}
\end{table*}



\bsp	
\label{lastpage}
\end{document}